\documentclass[journal=jacsat,manuscript=article]{achemso}

\usepackage[version=3]{mhchem} % Formula subscripts using \ce{}
\usepackage{amsfonts}
\author{Felix Eder}
\affiliation[Unige]
{University of Geneva, Department of Quantum Matter Physics, 24 Quai Ernest-Ansermet, CH-1211 Geneva, Switzerland}
\altaffiliation{These authors have contributed equally.}
\author{Catherine Witteveen}
\affiliation[Unige]
{University of Geneva, Department of Quantum Matter Physics, 24 Quai Ernest-Ansermet, CH-1211 Geneva, Switzerland}
\alsoaffiliation{University of Zurich, Department of Physics, 190 Winterthurerstr, CH-8057 Zurich, Switzerland}
\altaffiliation{These authors have contributed equally.}
\author{Enrico Giannini}
\affiliation[Unige]
{University of Geneva, Department of Quantum Matter Physics, 24 Quai Ernest-Ansermet, CH-1211 Geneva, Switzerland}
\author{Fabian O. von Rohr}
\affiliation[Unige]
{University of Geneva, Department of Quantum Matter Physics, 24 Quai Ernest-Ansermet, CH-1211 Geneva, Switzerland}
\email{FvR}

\title[]
  {Structural Modulation and Enhanced Magnetic Ordering in Incommensurate \ce{K_{1--\textit{x}}CrSe2} Crystals}

\abbreviations{AFM, FM, SXRD}

\begin{document}


\begin{tocentry}
\centering
\includegraphics[height=4cm]{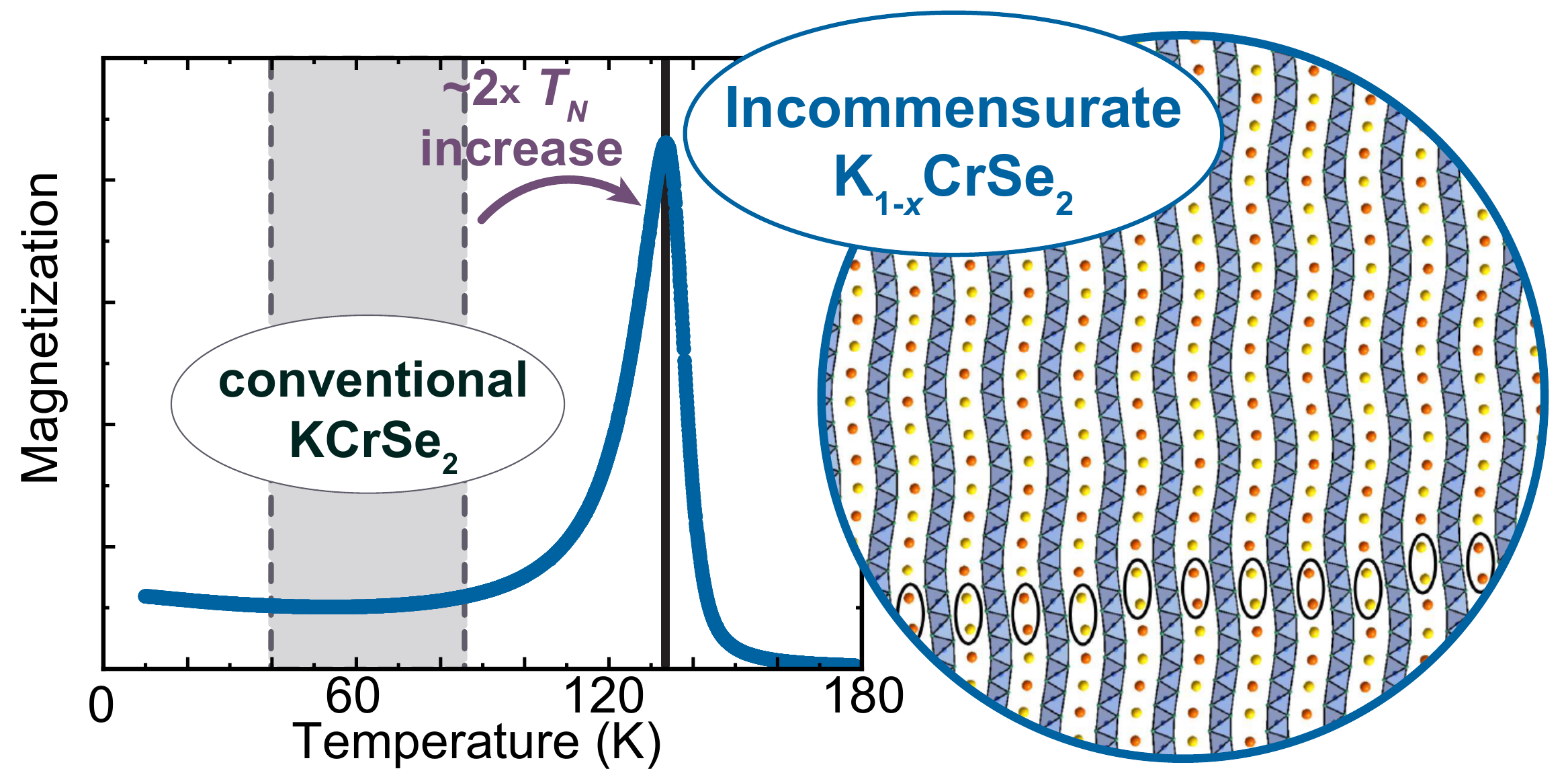}

\end{tocentry}

\begin{abstract}
Layered delafossite-type compounds and related transition metal dichalcogenides, characterized by their triangular net structures, serve as prototypical systems for exploring the intricate interplay between crystal structure and magnetic behavior. Herein, we report on the discovery of the compound K\textsubscript{1--\textit{x}}CrSe\textsubscript{2} (\textit{x} $\approx$ 0.13), an incommensurately modulated phase. Single crystals of this compound were grown for the first time using a K/Se self-flux. We find a monoclinic crystal structure with incommensurate modulation, that can be rationalized by a 3+1 dimensional model. This modulation compensates for the under-stoichiometry of K cations, creating pronounced undulations in the CrSe\textsubscript{2} layers. Our anisotropic magnetization measurements reveal that K\textsubscript{1--\textit{x}}CrSe\textsubscript{2} undergoes a transition to a long-range magnetically ordered state below $T_{\mathrm N}$ = 133 K, a temperature 1.6 to 3.3 times higher than in earlier reported \ce{KCrSe2} compounds. Our findings open new avenues for tuning the magnetic properties of these layered materials through structural modulation.
\end{abstract}

\section{Introduction}
Materials with triangular net structures and strong magnetic interactions have shown great potential for yielding exotic quantum states.\cite{chamorro2020chemistry,broholm2020quantum,kurumaji2019skyrmion} 
These structures host magnetically frustrated ground states and charge-ordering phenomena, and have even been associated with unconventional superconductivity, such as in water-intercalated Na\textsubscript{\textit{x}}CoO\textsubscript{2}.\cite{cheng2024superconductivity,weber2017trivalent,toth2016electromagnon,takada2003superconductivity,nocerino2023competition} Many of the prominently discussed triangular lattice compounds crystallize in structures related to the delafossite structure type.\cite{nocerino2023competition,kobayashi2019linear,moll2016evidence,zhang2024crystal} 

Delafossite-type chromium compounds, with the general formula \textit{A}Cr\textit{X}\textsubscript{2} (\textit{A} = +I oxidation state metal and \textit{X} = chalcogenide), consist of Cr\textit{X}\textsubscript{2} layers intercalated by the \textit{A} cations in a trigonal-rhombohedral stacking.\cite{marquardt2006crystal,Kobayashi2016} Deintercalation of the \textit{A} cations modifies the oxidation state from Cr\textsuperscript{III} to Cr\textsuperscript{IV}, resulting in materials with a van der Waals (vdW) gap, such as the series' end members \ce{CrTe2}, \ce{CrSe2}, and H\textsubscript{\textit{x}}\ce{CrS2}.\cite{peng2023recent,song2019soft,xian2022spin,roseler2025efficient}

For the top-down synthesis of these \ce{Cr\textit{X}_2} phases, the synthetic route has been the oxidative potassium deintercalation from \textit{A}Cr\textit{X}\textsubscript{2}.\cite{VanBruggen1980,Song2021} Specifically for CrSe\textsubscript{2}, the deintercalation of K cations from \ce{KCrSe2} removes them from the layer interspace, resulting in the vdW material as the final product. Full stoichiometric KCrSe\textsubscript{2} polymorphs have been prepared as polycrystalline material by solid-state synthesis from the elements.\cite{VanBruggen1980,Fang_1996} Previous reports have found that below a Néel temperature of \textit{T}\textsubscript{N} $\approx$ 85 K ($\mu_0 H =$ 20 mT) or \textit{T}\textsubscript{N} $\approx$ 40 K ($\mu_0 H =$ 860 mT), respectively, KCrSe\textsubscript{2} undergoes a transition to a long-range antiferromagnetic (AFM) state.\cite{Song2021,Wiegers1980, Fang_1996}

Calculations of the exchange coupling constants indicate that spins are ferromagnetically ordered within a layer plane and antiferromagnetically coupled to neighboring planes.\cite{Fang_1996} Similar magnetic behavior, with a comparable \textit{T}\textsubscript{N}, has been observed in the isotypic compounds KCrS\textsubscript{2} and NaCrSe\textsubscript{2}, as supported by neutron powder diffraction (NPD) data.\cite{vanlaar1973,engelsman1973}

Although the few existing studies on potassium chromium diselenides have primarily focused on the stoichiometric KCrSe\textsubscript{2} phase, some reports also explore compositions with lower potassium content.\cite{Wiegers1980,dijkstra1989,Fang_1996,VanBruggen1980} For one, an orthorhombic phase has also been proposed for a fast-cooled, slightly potassium-deficient phase, i.e., K\textsubscript{0.9}CrSe\textsubscript{2}.\cite{Fang_1996,Nikiforow1991} Second, for K contents of 0.6--0.8, a different type of rhombohedral stacking combined with smaller intralayer Cr---Cr distances was suggested. \cite{Wiegers1980} A recent study investigated the gradual deintercalation of KCrSe\textsubscript{2} using I\textsubscript{2}/acetonitrile and tracked the structural evolution from KCrSe\textsubscript{2} to more potassium-deficient phases like K\textsubscript{0.6--0.8}CrSe\textsubscript{2}.\cite{Song2021} All of these observed phases have been found to be structurally almost identical to stoichiometric \ce{KCrSe2} with a common pseudo-rhombohedral \textit{ABCABC} layer stacking typical for delafossite-type structures, and in the absence of any incommensurately modulated superstructure.

Mixed alkali-selenide or -sulfide fluxes are a potent tool for crystal growth and materials discovery, often capitalizing on the solubility of remaining \textit{A}/Se fluxes in organic solvents, most commonly N',N'-dimethyl formamide (DMF).\cite{Kanatzidis1990, Berseneva2023} The successful application of Li/Se fluxes, e.g. in the synthesis of EuSe\textsubscript{2},\cite{Aitken1998} and Na/Se fluxes, exemplified by the discovery of \ce{NaCu6Se4}\cite{Sturza2014} and \ce{NaCu4Se4}\cite{Chen2019}, has been demonstrated, and has been extended to K/Se fluxes as well with the example of KSmGeSe\textsubscript{4}.\cite{Martin2004}   
The synthetic path of the here reported K/Se flux synthesis is inspired by pure Se-based flux syntheses, exemplified by for the growth transition metal dichalcogenides \ce{PdSe2} and 2H-\ce{WSe2}.\cite{Gustafsson2018,Edelberg2019,oyedele2017pdse2} Instead of dissolution of remaining flux at room temperature in DMF, it is separated from the reaction products at high temperatures via high-temperature centrifugation. 
The viability of this method for mixed alkali-chalcogenide fluxes has been recently demonstrated in the crystal growth of LiCrTe\textsubscript{2}.\cite{witteveen2023synthesis} 

In this communication, we report on the discovery of the incommensurately modulated phase K\textsubscript{1--\textit{x}}CrSe\textsubscript{2} with \textit{x} $\approx$ 0.13. For this, we have developed a K/Se self-flux approach to obtain millimeter-sized single crystals of this compound. Our anisotropic magnetic measurements on oriented single crystals, parallel to the \textit{ab} plane and the \textbf{c*} axis, reveal rich magnetic properties that correspond to an A-type antiferromagnet (AFM) with its easy axis lying out of the $ab$ plane and a transition temperature of $T_{\mathrm N}$ = 133 K, which is 1.6 to 3.3 times higher than the values reported for the stoichiometric phase before. These findings expand on the structural and magnetic multifaceted nature of layered delafossite-type materials.

\section{Experimental}

\begin{table}
	\begin{center}
	\caption{Crystallographic data for K\textsubscript{1--\textit{x}}CrSe\textsubscript{2}}\label{tab1}
		\begin{tabular}{ll}	
\hline
Chemical formula 	&	 K\textsubscript{0.866}CrSe\textsubscript{2}	\\
Molar mass (g*mol$^{-1}$) 	&	 234.81	\\
$\rho$\textit{\textsubscript{calc.}} (g*cm$^{-3}$) & 4.467 \\
Temperature (K) & 100 \\
Space group 	&	 \textit{C}2/\textit{m}	\\
Super space group & \textit{C}2/\textit{m}(\textit{a}0\textit{g})00 \\
\textit{a} (\AA) & 6.3489(4) \\
\textit{b} (\AA) & 3.75881(14) \\
\textit{c} (\AA) & 8.2224(6) \\
\textit{$\beta$} (°) & 112.536(9) \\
\textit{V} (\AA$^3$) & 181.24(2) \\
Modulation vector \textbf{q} & 0.2675(4)\textbf{a*} -- 0.1585(4)\textbf{c*} \\
\textit{Z} & 2 \\
Crystal color & grey \\
Crystal shape & plate \\
Crystal size (mm$^3$) & 0.15$\times$0.05$\times$0.04 \\
Absorption correction & spherical \\
$\mu$ (mm\textsuperscript{--1}) & 23.96 \\
Diffractometer & Rigaku SuperNova \\
Radiation; $\lambda$ (\AA) & MoK\textsubscript{$\alpha$}; 0.71073 \\
Reflections used & 31627 \\
$\theta$\textsubscript{min} -- $\theta$\textsubscript{max} (°) & 2.5865 -- 30.5536 \\
\textit{h} range & --9 to 9 \\
\textit{k} range & --5 to 5 \\
\textit{l} range & --11 to 11 \\
\textit{m} range & --4 to 4 \\
Independent refl.\textsuperscript{a} & 2850/322/634/630/635/629 \\
Observed refl. (\textit{I} $<$ 3$\sigma$(\textit{I}))\textsuperscript{a} & 2317/302/594/576/499/346 \\
\textit{R\textsubscript{int}} & 0.058 \\
\textit{R1} (\textit{I} $<$ 3$\sigma$(\textit{I})); (\%)\textsuperscript{a} & 2.92/3.04/2.57/2.45/3.04/5.58 \\ 
\textit{wR2\textsubscript{all}} (\%) & 7.81 \\
Goodness of fit & 1.75 \\
CCDC deposition code 	&	 2381866	\\
\hline
	\end{tabular}
	\end{center}
\footnotesize{\textsf{[a] all reflections/main reflections/1\textsuperscript{st}/2\textsuperscript{nd}/3\textsuperscript{rd}/4\textsuperscript{th} order satellites }}
\end{table}

Single crystals of K\textsubscript{1--\textit{x}}CrSe\textsubscript{2} (\textit{x} $\approx$ 0.13) were prepared using a mixed K/Se self-flux.  
Synthesis conditions and sample handling were inspired by the successful growth of single crystals of LiCrTe\textsubscript{2} \cite{witteveen2023synthesis}. All handling of educts and reaction products was done in an argon-filled glovebox. 
Potassium (block, Sigma Aldrich, 99\%), chromium (powder, Alfa Aesar, 99.99\%), and selenium (pieces, Alfa Aesar, 99.999\%) were used as received and placed in the molar ratios 8:1:8 (total mass 1.5 g) in an alumina Canfield crucible set. 
This assembly, consisting of a bottom and top crucible separated by a frit-disc, was subsequently sealed in a quartz ampule under dynamic vacuum after having been purged three times with Ar \cite{Canfield2016}. An Ar partial pressure of 300 mbar was established in the ampule before sealing it.
The ampules were then heated in a muffle furnace (heating rate 30 °C/h) to 1000 °C, and slowly cooled to 750 °C over 96 hours (2.6 °C/h). 
The ampules were then removed from the furnace, flipped around, and immediately centrifuged to separate the crystals from the flux. Variation of the K-content in the flux (K:Cr:Se = \textit{n}:1:8; \textit{n} = 6--9) with the same synthetic procedure yields large crystals of K\textsubscript{1--\textit{x}}CrSe\textsubscript{2} as well.

The composition of grown crystals was analyzed using energy dispersive X-ray spectroscopy (EDS) on a JEOL JSM-7600F scanning electron microscope (SEM) with an accelerating voltage of 20 kV.
Powder X-ray diffraction (PXRD) measurements of the reaction product were performed on a Rigaku Smart-Lab diffractometer equipped with a 9 kW PhotonMax rotating anode Cu-source.
The crystal structure was investigated by single crystal X-ray diffraction (SXRD) on a Rigaku Supernova diffractometer using Mo-K\textsubscript{$\alpha$} radiation at 100 K. 
Unit-cell indexation, determination of the modulation vector, integration, and spherical absorption correction were performed with CrysalisPro.\cite{CrysAlisPRO2022}
After this, intensity data were imported into Jana 2020 \cite{Jana2020}, where the crystal structure was solved and subsequently refined against the main and satellite reflections up to fourth order. Details on the crystal structure determination are collated in Tab. \ref{tab1}.

The temperature-dependent and field-dependent magnetization was measured in a Physical Property Measurement System (Quantum Design PPMS DynaCool) equipped with a 9 T magnet with the vibrating sample magnetometer (VSM) option. The data was not corrected for the demagnetizing field. Further details on synthesis and sample characterization are collated in the SI.

\section{Results and discussion}

\subsection{Basic structure}

\begin{figure*}
\begin{center}
\includegraphics[width=16.5cm]{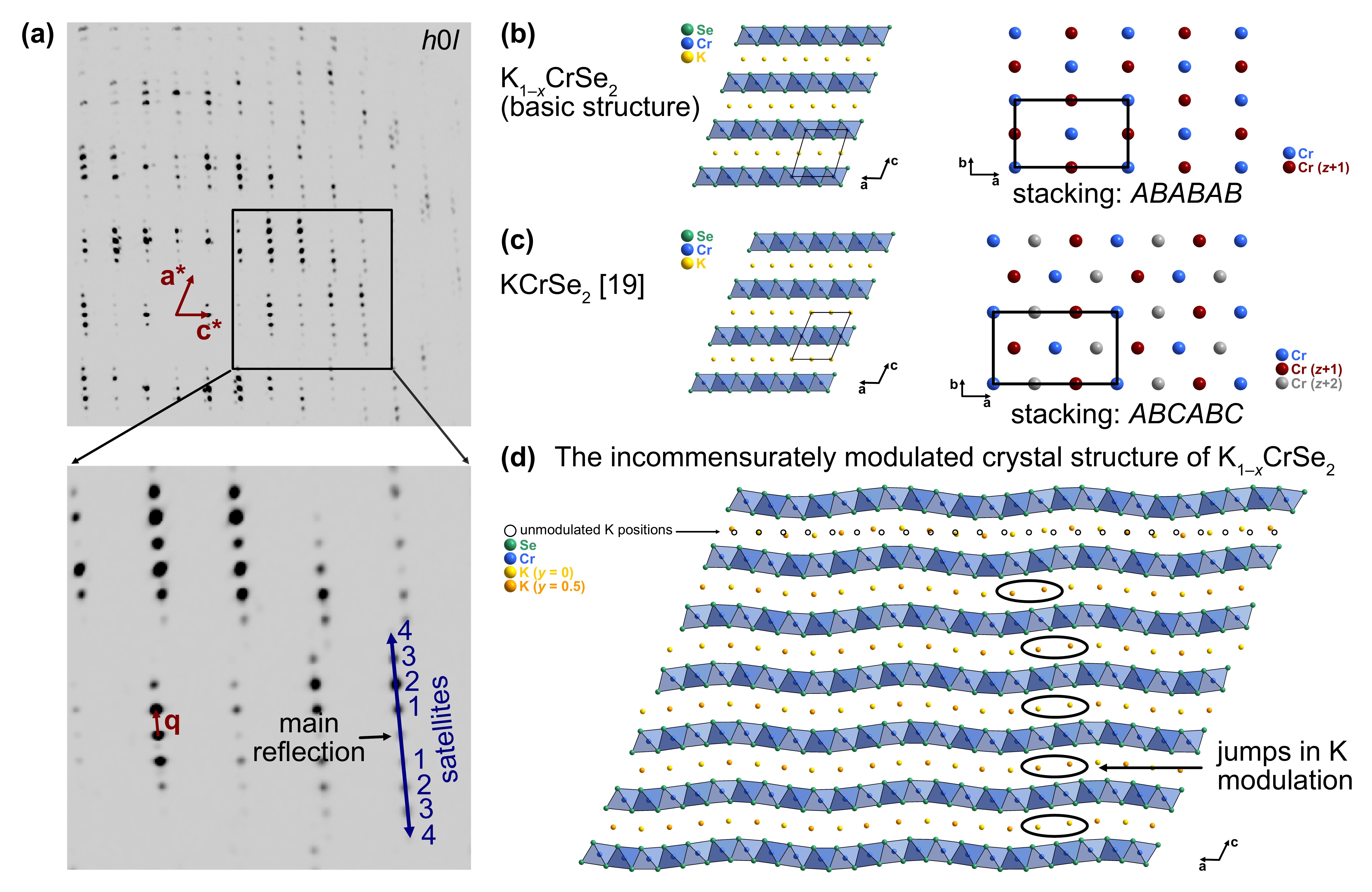}
\caption{Crystal Structure of K\textsubscript{1--\textit{x}}CrSe\textsubscript{2}: (a) Reconstructed \textit{h}0\textit{l} plane showing the satellite reflections of up to fourth order. (b) The basic structure of K\textsubscript{1--\textit{x}}CrSe\textsubscript{2} viewed along [010] (left) and the Cr atoms projected on the (001) plane, revealing the unconventional \textit{ABABAB} stacking. (c) The same projections for the crystal structure of KCrSe\textsubscript{2}\cite{Song2021}. (d) The crystal structure of K\textsubscript{1--\textit{x}}CrSe\textsubscript{2} viewed along [010]. An approximant structure corresponding to the size of 6$\times$1$\times$11 unit-cells of the basic structure is shown. K sites of the unmodulated basic structure are included at the top.}
\label{Fig1}
\end{center}
\end{figure*}

By application of the K/Se self-flux synthesis method, we have obtained plate-shaped crystals of K\textsubscript{1--\textit{x}}CrSe\textsubscript{2} with sizes up to 3 $\times$ 2 $\times$ 0.3 mm\textsuperscript{3}. The crystal platelets have elongated shapes and a metallic silvery luster. We find that the crystals are sensitive to air/humidity, and elongated exposure to atmospheric conditions leads to the formation of round stains on the surface (see Fig. S1) as a first degradation sign.
The PXRD pattern of the reaction product (see Fig. S2) shows almost phase-pure K\textsubscript{1--\textit{x}}CrSe\textsubscript{2}, including pronounced satellite reflections, with a small impurity of the \ce{K2Se} flux.

In Fig. \ref{Fig1}(a), we show the reconstructed \textit{h}0\textit{l} plane from our SXRD measurements of K\textsubscript{1--\textit{x}}CrSe\textsubscript{2} displaying pronounced satellite reflections. Based on only the main reflections, we can determine the basic structure in the monoclinic space group \textit{C}2/\textit{m} with the lattice parameters \textit{a} = 6.3489(4) \AA, \textit{b} = 3.75881(14) \AA,  \textit{c} = 8.2224(6) \AA, and \textit{$\beta$} = 112.536(9)° leading to a cell volume of \textit{V} = 181.24(2) \AA $^3$ . The basic structure of K\textsubscript{1--\textit{x}}CrSe\textsubscript{2} is depicted in Fig. \ref{Fig1}(b) and comprises three crystallographic sites, one occupied by Cr (multiplicity 2, Wyckoff letter \textit{d}, site symmetry 2/\textit{m}), Se (4 \textit{i}, \textit{m}) and K (2 \textit{a}, 2/\textit{m}) atoms respectively. The crystal structure is formed by CrSe\textsubscript{2} layers oriented parallel to (001). 
They consist of slightly distorted CrSe\textsubscript{6} octahedra, which are connected to six neighboring units by edge-sharing. By this, the Cr atoms are arranged in a triangular substructure. The K cations are located in the interlayer space and are coordinated by the Se atoms of two neighboring layers.

The basic structure of K\textsubscript{1--\textit{x}}CrSe\textsubscript{2} and the monoclinic unit-cell of KCrSe\textsubscript{2}\cite{Song2021} (Fig. \ref{Fig1}(c)) appear only at first glance very similar. In KCrSe\textsubscript{2}, the \textit{a}/\textit{b} ratio of 1.7322 is almost identical to $\sqrt{3}$ and underlines the close relationship to the rhombohedral \textit{R}$\Bar{3}$\textit{m} space group originally assigned \cite{Wiegers1980,Fang_1996}. 
However, in the here reported K\textsubscript{1--\textit{x}}CrSe\textsubscript{2}, the metrics of the unit-cell are heavily distorted (\textit{a}/\textit{b} = 1.689). Another difference concerns the stacking of the layers when viewed perpendicular to the layer plane along the \textbf{c*} direction. While previous KCrSe\textsubscript{2} phases exhibit an \textit{ABCABC} stacking of layers, the stacking in our reported K\textsubscript{1--\textit{x}}CrSe\textsubscript{2} phase corresponds to an unusual \textit{ABABAB} stacking. This becomes apparent when comparing the respective structures in the side views of Fig. \ref{Fig1}(b)\&(c). This type of stacking has not been observed in any of the other \textit{A}Cr\textit{X}\textsubscript{2} compounds. In general, no relatable compound is found, when searching for the unit-cell parameters of the basic structure in the current version of the ICSD \cite{Zagorac2019} with a tolerance of up to 7 \%.
The closest relation of the structural features can be found with Na\textsubscript{0.78}CoO\textsubscript{2}, for which a similar modulation behavior including the same super space group, suggested from neutron powder diffraction data, was reported recently.\cite{Miyazaki2021}

\subsection{Incommensurate modulation}

The satellite reflections, which dominate the diffraction pattern of K\textsubscript{1--\textit{x}}CrSe\textsubscript{2}, are visible up to fourth order in the reconstructed reciprocal \textit{hnl} ($n \in \mathbb{Z}$) planes as presented in Fig. \ref{Fig1}(a), but not in the \textit{hkn} and \textit{nkl} ($n \in \mathbb{Z}$) planes (see Fig. S3).
All observed reflections, including satellites, can be explained under consideration of one modulation vector \textbf{q}. \cite{TableC} 
\begin{equation} \label {eq1}
			\textbf{q} = q_1\textbf{a*}+q_2\textbf{b*}+q_3\textbf{c*}
\end{equation} 

The modulation vector of the crystal used for the presented structure determination amounts to \textit{q\textsubscript{1}} = 0.2675(4), \textit{q\textsubscript{2}} = 0, and \textit{q\textsubscript{3}} = --0.1585(4).
We find that the modulation vector does not change within its standard deviation when heating the crystal in 10--20 K steps from 90--350 K (see Fig. S4). 
However, we observe some slight variation in its size when investigating different crystals from the same synthesis batch: 0.2578(3) $<$ \textit{q}\textsubscript{1} $<$ 0.2689(5), and --0.1529(5) $>$ \textit{q}\textsubscript{3} $>$ --0.1598(5).

Based on the main reflections and satellites up to fourth order, the crystal structure of K\textsubscript{1--\textit{x}}CrSe\textsubscript{2} was solved and refined in the 3+1 dimensional super space group \textit{C}2/\textit{m}(\textit{a}0\textit{g})00 (details in the SI).
A large sector of the crystal structure of K\textsubscript{1--\textit{x}}CrSe\textsubscript{2} is depicted in Fig. \ref{Fig1}(d). 
The most notable difference to the other \textit{AMX}\textsubscript{2} crystal structures (Fig. \ref{Fig1}(b)$\And$(c)) is a pronounced undulation of the CrSe\textsubscript{2} layers propagating along \textbf{a}. 
This undulation concerns all three atom types, as is clearly visible in the differences of the atomic positions in the modulated structure and the basic structure along the \textbf{c} direction (see Fig. S8(a)).

\begin{figure*}
\begin{center}
\includegraphics[width=16.5cm]{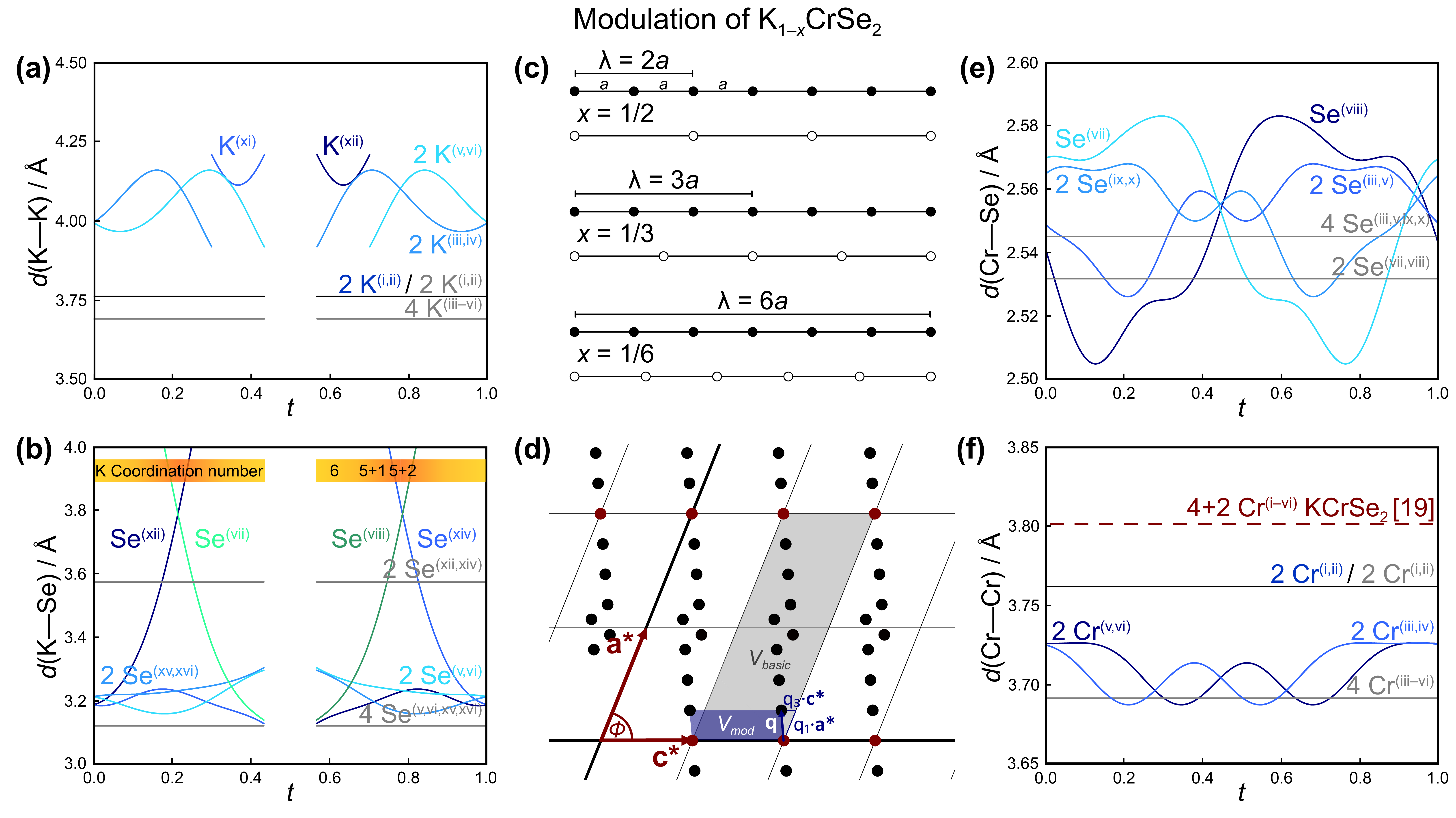}
\caption{Modulation of K\textsubscript{1--\textit{x}}CrSe\textsubscript{2}: (a), (b) Modulation of K---K and K---Se distances. Modulated distances are drawn colored, distances unaffected by the modulation ($\pm$ \textbf{b}) black, and distances in the basic structure gray. Symmetry codes are given in the SI: section 2G. (c) Dependence of the translational period of a 1D composite system on the relative occupations of the subsystems. (d) Schematic view of the satellite reflections in the \textit{h}0\textit{l} plane. Main reflections are drawn red, satellite reflections black, and the reciprocal volumes spanned by the main reflections or all reflections in gray and blue, respectively. (e), (f) Modulation of Cr---Se and Cr---Cr distances. Same colors as in (a), (b) apply. Distances in full-stoichiometric \ce{KCrSe2}\cite{Song2021} are shown in red.}
\label{Fig2}
\end{center}
\end{figure*}

We find that the reduced K content is the root cause of the observed incommensurate modulation. Instead of following the basic structure with occasional K vacancies, the K atoms attempt to equally distribute in the interlayer space, enlarging the K---K distances (Fig. \ref{Fig2}(a); black circles at the top of Fig. \ref{Fig1}(d)). 
By this, the K cations exhibit larger K---K distances (3.924(5)--4.204(6) \AA) relative to comparable Cr---Cr distances (3.685(3)--3.725(3) \AA) when progressing along \textbf{a}. 
These different translational periods create a mismatch in the crystal structure, which is visible in the distribution of the K cations.
A perfect equidistant spacing of the K cations is hindered by the presence of the Se atoms of adjacent layers, which coordinate the K cations. 
Hereby, the K cations show varying shapes of coordination numbers (CNs) and geometry. 
In most cases, a CN of 6 is realized in either a trigonal prismatic or antiprismatic coordination, while for some K cations the CN is better defined as 5+1 or 5+2 (Fig. \ref{Fig2}(b)).
The undulation of the (001) oriented layers is then the result of the crystal structure of K\textsubscript{1--\textit{x}}CrSe\textsubscript{2} attempting to compensate this mismatch between the CrSe\textsubscript{2} and K\textsubscript{1--\textit{x}} layers.
In contrast, the interatomic distances in the \textbf{b} direction are entirely unaffected by the modulation.  

Based on the \textit{C}-centered basic structure, neighboring Cr atoms are displaced by \textbf{b}/2, when progressing along \textbf{a}.
For the K cations, this rule is periodically broken, leading to the presence of K pairs at the same \textit{y} coordinate (right of Fig. \ref{Fig1}(d)), further highlighting the mismatch between K cations and CrSe\textsubscript{2} layers. 
These K pairs at the same \textit{y} values occur every 6 to 7 K cations (accurately 2/\textit{q\textsubscript{1}}--1 $\approx$ 6.477).

The reduced K content in K\textsubscript{1--\textit{x}}CrSe\textsubscript{2} is not only the origin of the incommensurate modulation, it is also directly connected to the size of the modulation vector, and can, vice-versa, be determined from it.
In a simple one-dimensional case of two subsystems with different translational periods (more details see SI: section 2E), caused by an under-occupation of the second system by a factor of 1--\textit{x}, the translational period of the whole system $\lambda$ is equal to 1/\textit{x} for \textit{x} $\leq$ 0.5 (illustrated in Fig. \ref{Fig2}(c)). 
As $\lambda$ is connected to the size of the modulation vector in this direction \textit{q} by $\lambda$ = 1/\textit{q}, this leads to the simple relation \textit{x} = \textit{q}. 
Due to the \textit{C}-centering of the unit-cell of K\textsubscript{1--\textit{x}}CrSe\textsubscript{2} and therefore the presence of two CrSe\textsubscript{2} units and K cations per lattice vector \textbf{a}, the relation changes to \textit{x} = \textit{q}\textsubscript{1}/2 in our case.

As the modulation vector \textbf{q} = 0.2675(4)\textbf{a*} -- 0.1585(4)\textbf{c*} is not parallel to one of the three reciprocal axes, we can generalize the previous approach by comparing the reciprocal volumina spanned by just the main reflections (basic structure) or including the satellites (modulated structure) (Fig. \ref{Fig2}(d)).
As \textbf{b*} is unaffected by the modulation, it suffices to compare the colored areas depicted.
It is not necessary to know the size of the other reciprocal axes \textit{a*} and \textit{c*} or the angle $\phi$ between them, as they all divide out of eq. \ref{eq5}.

\begin{equation} \label{eq5}
x=\frac{V_{mod}}{V_{basic}}=\frac{q_1*|\textbf{a*}|*|\textbf{b*}|*|\textbf{c*}|*sin(\phi)}{2*|\textbf{a*}|*|\textbf{b*}|*|\textbf{c*}|*sin(\phi)}=\frac{q_1}{2}
\end{equation} 
Based on \textit{q}\textsubscript{1} = 0.2675(4), this leads to \textit{x} = 0.1338(2) and, henceforth, a sum formula of K\textsubscript{0.8662(2)}CrSe\textsubscript{2}.

The relation between \textit{q\textsubscript{1}} and \textit{x} can be supported further by varying the $\Delta$ parameter of the Crenel function used to represent the understoichiometry of the K cations in the refinement (see Fig. S8(c) \& (d)). 
The temperature-independence (90--350 K) of the modulation vector and the fact that slightly varying modulation vectors (SXRD: \textit{q}\textsubscript{1} = 0.2578(3)--0.2689(3); PXRD: \textit{q}\textsubscript{1} = 0.2732(4)) were observed for different crystals of the same batch are likewise in line with this model. 
We find the small compositional variability on different crystals not surprising, as we likewise found small variations of the K content (0.83(2)--0.87(2)) when we checked the composition of twelve different crystals from the synthetic batch by EDS (details see SI). A larger range in modulation vectors can be observed when analyzing single crystals from synthetic batches with different K-contents. On crystals investigated, the range of modulation vectors is 0.2578(3) $<$ \textit{q}\textsubscript{1} $<$ 0.3404(10) and --0.1529(5) $>$ \textit{q}\textsubscript{3} $>$ --0.1954(14). This indicates that K\textsubscript{1--\textit{x}}CrSe\textsubscript{2} grows with a certain homogeneity range of \textit{x} and not as a line compound.

While the \textit{q}\textsubscript{1} component of the modulation vector (0.2675(4)) is directly linked to the reduced K content, \textit{q}\textsubscript{3} (--0.1585(4)) describes the evolution of the stacking. 
The undulated CrSe\textsubscript{2} sheets are stacked almost perfectly along \textbf{c*}, a slightly less negative \textit{q}\textsubscript{3} component of --0.1328 would correspond to this (see SI: section 2F). 
The difference between \textit{q}\textsubscript{3} and this ideal value describes a small shift along \textbf{a} of the undulations of neighboring layers.
For the presented crystal structure, this means that after approximately 39 layers, the undulation has shifted by one of its translational periods $\lambda$ in the \textbf{a} direction (see Fig. S10).

While the modulation originates from the distribution of K cations, the CrSe\textsubscript{2} layers are strongly influenced by it. 
Besides their pronounced undulation, which has a translational period $\lambda$ of 1/\textit{q\textsubscript{1}} = 1/0.2673 $\approx$ 3.741 \textbf{a} vectors of the basic structure, the Cr---Se distances vary in the range of 2.504(3)--2.582(3) \AA. This is a much larger interval than in the basic structure, as shown in Fig. \ref{Fig2}(e). As a result, the bond valence sum (BVS)\cite{brownBVS} of the Cr atoms exhibits a variance of more than 0.1 valence units (v.u.) as well (see Fig. S11(a)). 
Likewise, the Cr---Cr distances along the $<$110$>$ directions (3.685(3)--3.725(3) \AA) (Fig. \ref{Fig2}(f)) and the Cr---Se---Cr angles (94.08(6)--96.12(6)° for $<$010$>$ Cr---Cr pairs and 91.40(10)--94.90(10)° for those along $<$110$>$; see Fig. S11(b)) are varied by the modulation. 
In general, the Cr---Se---Cr angles are slightly lower than the ones reported for KCrSe\textsubscript{2} \cite{Song2021} (95.24°).
All of this underlines that not all Cr atoms in K\textsubscript{1--\textit{x}}CrSe\textsubscript{2} have an identical bonding environment, which can influence and complicate their respective magnetic responses and couplings.
Furthermore, we find that modulation and the resulting undulation of the CrSe\textsubscript{2} layers also influence the stacking sequence in general. 
The unusual \textit{ABABAB} stacking allows a placement of the undulated sheets without significant translation along \textbf{a} on top of each other.

\subsection{Magnetic properties}

\begin{figure*}
\begin{center}
\includegraphics[width=16.5cm]{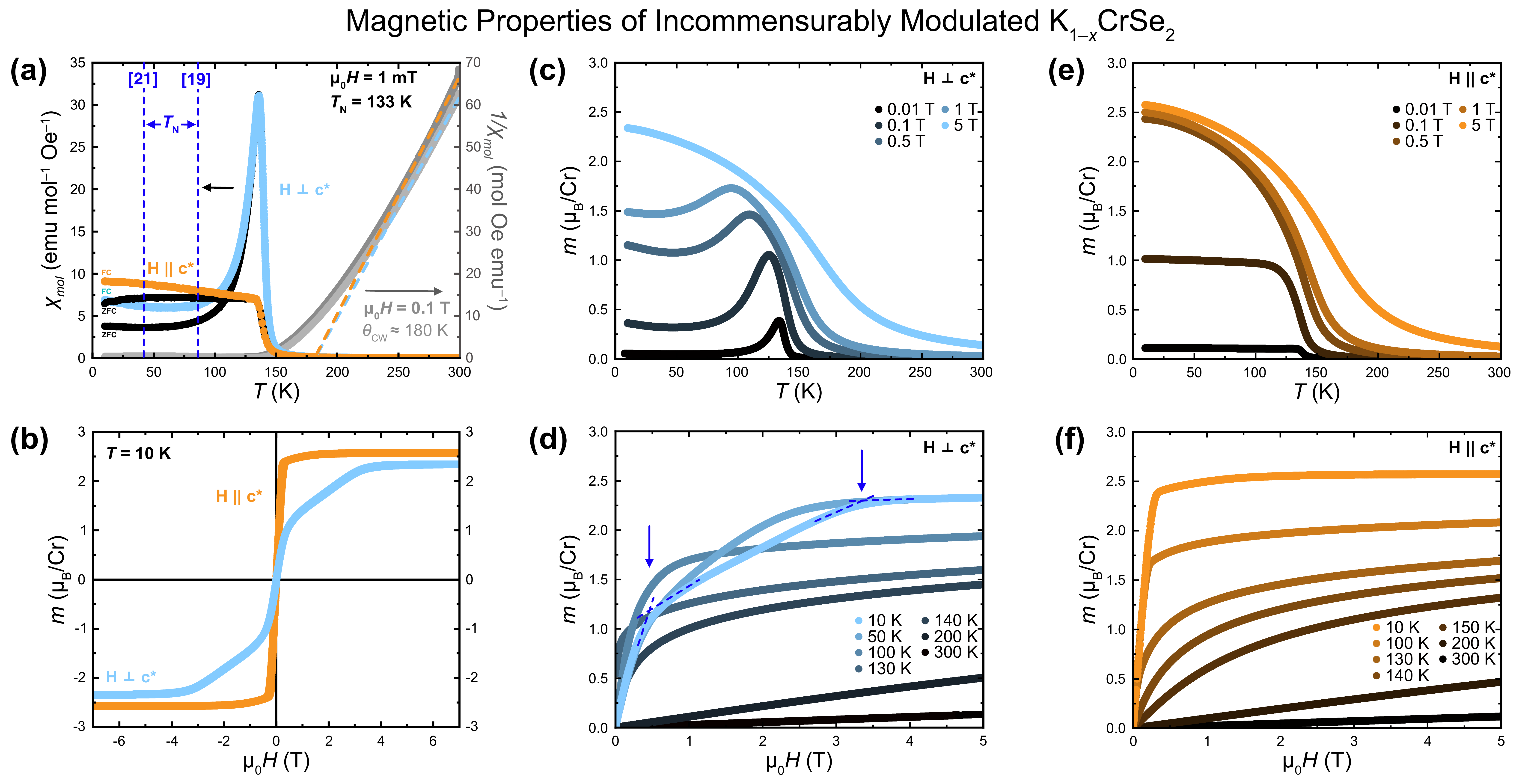}
\caption{Magnetic properties of K\textsubscript{1--\textit{x}}CrSe\textsubscript{2}: (a) ZFC (black) and FC (colored) molar susceptibility for both orientations of K\textsubscript{1--\textit{x}}CrSe\textsubscript{2} at $\mu_0 H =$ 1 mT (left axis). The inverse susceptibility of both orientations is taken at $\mu_0 H =$ 0.1 T (right axis). The previously reported \textit{T\textsubscript{N}} values for the full-stoichiometric powder samples of \ce{KCrSe2} are shown as a blue dotted line for comparison. It should be noted that Song \textit{et al.}\cite{Song2021} have speculated that the $T_{\rm N}$ in the full-stoichiometric phase is strongly field-dependent. (b) Field-dependent magnetic moment for both orientations taken at 10 K. (c), (e) Temperature-dependent magnetic moment per chromium atom for both orientations, with an observed metamagnetic transition when the field is applied perpendicular to \textbf{c*}. (d), (f) Field-dependent magnetic moment per chromium atom for both orientations. The blue arrows indicate the change of slope in the curve taken at 10 K.}
\label{Fig3}
\end{center}
\end{figure*}

In Fig. \ref{Fig3}, we show the temperature- and field-dependent magnetic properties of K\textsubscript{1--\textit{x}}CrSe\textsubscript{2} single crystals. The availability of single crystals enabled us to perform direction-dependent magnetization measurements on oriented crystal plates. The magnetic field was applied parallel to the crystallographic \textbf{a}, \textbf{b}, and \textbf{c*} axes. Hereby, the measurements parallel to the layer plane yielded qualitatively the same responses (\textbf{H} $\perp$ \textbf{c*}, see Fig. S12)

The temperature-dependent molar magnetic susceptibility, defined here as $\chi_{mol}$ = M(mol)/H, shown in Fig. \ref{Fig3}(a) reveals the distinct transition to a long-range magnetically ordered state at $T_{\mathrm N}$ = 133 K. It is worth highlighting that this transition temperature measured here is substantially higher (factor 1.6 at similar fields) than the ordering temperature in earlier reports of stoichiometric \ce{KCrSe2} shown as blue dotted line in Fig. \ref{Fig3}(a) for reference.\cite{Song2021} This increase is even more remarkable since, while incommensurate magnetic ordering based on ordered nuclear structures are quite common, structural incommensurate modulation is usually not expected to be favorable to the establishment of long-range magnetic ordering.

We find significant differences in the magnetic properties of these crystals depending on whether the field is applied perpendicular or parallel to the \textbf{c*} direction: For the \textbf{H} $\perp$ \textbf{c*} orientation, a sharp peak is observed at the transition temperature, below which the magnetic moment rapidly decreases, characteristic of an antiferromagnetic (AFM) long-range order. In the \textbf{H} $\parallel$ \textbf{c*} orientation, the magnetic moment saturates to an almost temperature-independent value below the transition, suggesting a ferromagnetic (FM) order (Fig. \ref{Fig3}(a), left axis). 

The inverse susceptibility results yield positive Curie-Weiss temperatures (fitted between 250 K -- 300 K), $\Theta_{CW} \approx 180$ K, for both crystallographic directions (Fig. \ref{Fig3}(a), right axis). Slightly lower but similar positive Curie-Weiss temperatures have been previously reported for powder samples of KCrSe\textsubscript{2}, albeit the lower transition temperatures.\cite{Wiegers1980, Fang_1996, Song2021} These positive $\Theta_{CW}$ correspond to nearest-neighbor magnetic interactions that are predominantly FM. This is in line with the Goodenough-Kanamori-Anderson (GKA) rules, which allow us to predict the type of coupling between the Cr ions as a function of the geometry of the structure. In the Cr\textsuperscript{III} (3\textit{d}\textsuperscript{3}) triangular lattice of 2D materials, there exists a competition between the AFM direct Cr---Cr coupling and the superexchange coupling. The latter is FM when the Cr---\textit{X}---Cr form a 90° angle. The smaller the Cr---Cr distance, the more influential the first contribution becomes. This dependence on the nearest-neighbor distances has been shown exemplarily for intercalated Cr\textit{S}\textsubscript{X} phases and explains the complex AFM helimagnetic ordering observed in the intercalated phase LiCrSe\textsubscript{2}.\cite{nocerino2023competition} 

As shown in Fig. S11(b), the Cr---Se---Cr angles of our modulated compound are closer to 90° than the angles reported for KCrSe\textsubscript{2}, leading to stronger superexchange couplings albeit shorter Cr---Cr distances. This explains the higher Curie-Weiss temperature compared to the full-stoichiometric sample.\cite{Song2021} 

In Fig. \ref{Fig3}(b) we show the field-dependent magnetization in 5 quadrants for both orientations taken at 10 K. 
In Fig. \ref{Fig3}(c)\&(d), are field- and temperature-dependent magnetization measurements with the field applied perpendicular to \textbf{c*}. A characteristic metamagnetic transition from an AFM to an FM ordered state is observed. 
In Fig. \ref{Fig3}(e)\&(f), we present the field- and temperature-dependent magnetization measurements in the vicinity of the magnetic transition with the field applied $\parallel$ to \textbf{c*}. There, the magnetic moment remains saturated below $T_{\mathrm N}$ across all fields.

The different magnetic behavior can be well described by having a closer look at the exemplary $m$($H$) at $T$ = 10 K in Fig. \ref{Fig3}(b) and Fig. \ref{Fig3}(e)\&(f) :

(i) In the \textbf{H} $\perp$ to \textbf{c*} orientation, the $m$($H$) rises fast with increasing fields for very low fields up to $\mu_0 H$ = 0.4 T (blue arrow). For fields from 0.4 T on, we observe a much less steep slope until a magnetic saturation of 2.40 $\mu$\textsubscript{B} is reached at $\mu$\textsubscript{0}\textit{H} = 3.25 T (blue arrow). (ii) In the \textbf{H} $\parallel$ \textbf{c*} orientation, the $m$($H$) rises fast with increasing fields to reach a slightly higher saturation value of 2.57 $\mu$\textsubscript{B} (closer to the expected value of 2.87 $\mu$\textsubscript{B}) for a field of $\mu_0 H$ = 0.3 T, revealing that the easy axis is along the \textbf{ c *} direction.

These observations can be rationalized by the occurrence of long-range A-type antiferromagnetic (AFM) order below $T_{\mathrm{N}}$ = 133 K. In this configuration, the magnetic moments are oriented perpendicular to the $ab$ plane and are ferromagnetically coupled within that plane, while the FM planes are antiferromagnetically coupled between layers. Thereby, the AFM interlayer interaction appears to be weak and the FM state to be only slightly higher in energy than the AFM order. This allows for a swift transition to the FM state for small external fields when applied parallel to the easy axis and for a broad metamagnetic transition when applied perpendicular to the easy axis. Specifically, when the field is applied perpendicular to the easy axis, the external field forces the magnetic moments out of their preferred orientation. 

However, a more complex magnetic scenario, such as, e.g., a ferrimagnetic order, could be present in \ce{K_{1--\textit{x}}CrSe2}, potentially stabilized by the incommensurate structural modulation, and the resulting differing Cr environments. The elucidation of the true magnetic ground state therefore will require future investigation with specialized probes.

\section{Conclusion}

By application of a K/Se self-flux synthesis method, we have obtained large single crystals of K\textsubscript{1--\textit{x}}CrSe\textsubscript{2} (\textit{x} $\approx$ 0.13). Our synthesis underlines the viability of \textit{A}/Se-fluxes paired with high-temperature centrifugation for single crystal growth.

We find a monoclinic crystal structure with incommensurate modulation, leading to a 3+1 dimensional model for this compound. The substoichiometry of the K cations causes an incommensurate modulation, which manifests, due to the coordination of the K cations by the Se atoms of neighboring CrSe\textsubscript{2} layers, in pronounced undulation of the whole layers. 

We find the wave-like shape of the CrSe\textsubscript{2} sheets to be the reason for their unique stacking, which has not been observed in other \textit{AMX}\textsubscript{2} compounds. Likewise, the distortion of the lattice parameters with \textit{a} $<<$ $\sqrt{3}$\textit{b} is another consequence of the undulation of the layers.

Our magnetization measurements reveal that K\textsubscript{1--\textit{x}}CrSe\textsubscript{2} likely undergoes a transition to an A-type antiferromagnetic order below $T_{\mathrm N}$ = 133 K, where the spins are aligned ferromagnetically in the ab-plane and the planes are coupled antiferromagnetically. The easy axis is oriented along \textbf{c*}, as can be deduced from our field-dependent magnetization measurements. At low fields applied perpendicular to the easy axis, we observe an antiferromagnetic order, which gradually shifts to ferromagnetic order upon increasing the field strength, characteristic for a metamagnetic transition. A saturation of 2.4 $\mu$\textsubscript{B} is reached at 3.25 T.
When the field is applied parallel to the easy axis, the system only shows a ferromagnetic order with a lower saturation field of 0.3 T but a higher saturation moment of 2.57 $\mu$\textsubscript{B} .

These magnetization measurements, henceforth, suggest the presence of rich and complex magnetic interactions in this system. This is supported by the observation of a strongly enhanced transition temperature (in comparison with stoichiometric \ce{KCrSe2} phases), the competition between AFM and FM orders, as well as the remarkably broad metamagnetic transition along the hard axis. In addition to that, the structural modulation associated with K off-stoichiometry, which influences the Cr---Se---Cr angles and thereby the Cr---Cr superexchange interaction, is likely to play a role in stabilizing a more intricate magnetic ground and dynamic states.

The here presented results on this hitherto unreported K\textsubscript{1--\textit{x}}CrSe\textsubscript{2} phase highlights the magnetic complexity resulting from the mixed oxidation of the chromium atoms in a two-dimensional nature of the structure, paired with the geometric arrangement of the K cations leading to an undulation of the layers. These findings open up new avenues in the field of the structure-property relations of layered magnetic materials.

\begin{acknowledgement}

The authors thank Berthold Stoeger for helpful discussions on the incommensurate modulation. The help of Stefano Gariglio in the setup and performance of the PXRD measurements is thankfully acknowledged. This work was supported  by the Swiss National Science Foundation under Grants No. PCEFP2\_194183 and No. 200021-204065.

\end{acknowledgement}

\begin{suppinfo}

Further details on sample preparation, characterization and more insights on the modulated crystal structure.

\end{suppinfo}

\bibliography{KxCrSe2}

\end{document}